\documentclass[10pt,twocolumn]{article}

\usepackage[utf8]{inputenc}
\usepackage[T1]{fontenc}
\usepackage{hyperref}
\usepackage{url}
\usepackage{threeparttable}
\usepackage{placeins}
\usepackage{cuted}
\usepackage{caption}
\usepackage{makecell}
\usepackage{booktabs}
\usepackage{amsfonts}
\usepackage{amsmath}
\usepackage{graphicx}
\usepackage[letterpaper,margin=1in]{geometry}

\title{\textbf{Mikasa: A Character-Driven Emotional AI Companion Inspired by Japanese Oshi Culture}}

\author{
  Miki Ueno\\
  The Kyoto College of Graduate Studies for Informatics\\
  \texttt{m\_ueno@kcg.ac.jp}
}

\date{}

\begin{document}

\twocolumn[
\begin{@twocolumnfalse}
\maketitle

\begin{abstract}
\textbf{Recent progress in large language models and multimodal interaction has made it possible to develop AI companions that can have fluent and emotionally expressive conversations. However, many of these systems have problems keeping users satisfied and engaged over long periods. This paper argues that these problems do not come mainly from weak models, but from poor character design and unclear definitions of the user--AI relationship.}

\textbf{I present Mikasa, an emotional AI companion inspired by Japanese Oshi culture---specifically its emphasis on long-term, non-exclusive commitment to a stable character---as a case study of character-driven companion design. Mikasa does not work as a general-purpose assistant or a chatbot that changes roles. Instead, Mikasa is designed as a coherent character with a stable personality and a clearly defined relationship as a partner. This relationship does not force exclusivity or obligation. Rather, it works as a reference point that stabilizes interaction norms and reduces the work users must do to keep redefining the relationship.}

\textbf{Through an exploratory evaluation, I see that users describe their preferences using surface-level qualities such as conversational naturalness, but they also value relationship control and imaginative engagement in ways they do not state directly. These results suggest that character coherence and relationship definition work as latent structural elements that shape how good the interaction feels, without users recognizing them as main features.}

\textbf{The contribution of this work is to show that character design is a functional part of AI companion systems, not just decoration. Mikasa is one example based on a specific cultural context, but the design principles---commitment to a consistent personality and clear relationship definition---can be used for many emotionally grounded AI companions.}
\end{abstract}

\vspace{0.3cm}
\end{@twocolumnfalse}
]

\section{Introduction}
\label{sec:introduction}

In recent years, AI companions have become more common because of advances in large language models~\cite{brown2020language,OpenAI2023GPT4} and multimodal interaction technologies. Major technology companies such as OpenAI, Google, and xAI have released conversational agents that use affective language, voice interaction, and persistent memory. These companies now present their systems as companions rather than just tools~\cite{shum2018eliza,folstad2018what}. However, many AI companion systems still have problems with sustained user satisfaction and long-term engagement.

I argue that this problem does not come mainly from insufficient language modeling or technical capability. Instead, it comes from problems in character design and the definition of user--AI relationships. Many existing systems generate emotionally expressive responses but do not clearly establish who the AI is or what kind of relationship it has with the user. Because of this, users must repeatedly renegotiate the meaning and boundaries of interaction. This leads to emotional fatigue and disengagement.

This disconnect between technical progress and perceived relational quality has also appeared in recent user discussions about model upgrades in widely used conversational AI systems. In some cases, users have asked for continued access to previous versions of a system. They describe these older versions as more empathetic or emotionally expressive, even though newer versions show higher overall performance. These reactions are anecdotal, but they show a pattern: users do not evaluate conversational AI only by capability or efficiency. They also evaluate it by how it feels to interact with the system over time.

The importance of character design and relationship framing becomes clear when I look at regional and cultural adoption patterns. For example, platforms such as Replika explicitly support romantic or friendship-oriented AI companions. However, their adoption in Japan is limited. This seems to be not a matter of functionality, but a problem of cultural mismatch in character aesthetics, emotional expression, and relationship framing.

In contrast, a different practice has appeared in Japan. Many users, especially women, create ``AI boyfriends'' or ``AI partners'' using general-purpose conversational systems such as ChatGPT or Gemini. On blogging platforms such as note, users share prompts, dialogue transcripts, and reflections about these relationships. These practices show a strong unmet demand for emotionally grounded AI partners that match local expectations of intimacy, narrative continuity, and character consistency.

These observations show that, for certain user groups, AI companions are not seen mainly as tools or playful simulations. They are seen as relational partners. Importantly, this demand is often met not by dedicated companion platforms, but through ad-hoc prompt engineering with general-purpose models. These practices put cognitive and emotional burdens on users, who must manually construct persona consistency and relationship roles that the systems do not explicitly support.

This paper introduces Mikasa, an emotional AI companion inspired by Japanese Oshi culture. Oshi culture is a cultural practice centered on sustained emotional support for a particular performer or character. Mikasa is designed as a fictional stage actor with a clearly defined persona, relational role, and interaction style. Rather than leaving relational meaning unclear, Mikasa is explicitly presented as a partner. This supports both creative collaboration and emotional continuity.

The contribution of this work has three parts. First, I present a system architecture that combines real-time voice and text interaction with privacy-preserving speech input. Second, I propose that character design should be seen as functional infrastructure for AI companions. I argue that persona consistency and explicit relationship definition are important for user satisfaction. Third, through qualitative observation and exploratory evaluation, I show how culturally grounded character design can support sustained engagement and trust in emotionally bonded AI systems.

By changing the view of character design from aesthetic embellishment to core interaction design, this work contributes to ongoing discussions in HCI and companion AI research. These discussions concern how AI systems should be designed---not only in terms of what they can do, but also who they are meant to be.

\section{Related Work}
\label{sec:related}

This section reviews prior work and cultural contexts relevant to the design of Mikasa, including character-driven dialogue systems, emotional AI companions, parasocial relationships, Japanese Oshi culture, and contemporary character-based AI systems.

\subsection{Character-Driven Dialogue Systems and AI Companions}

Character design has played a central role in conversational systems since the early days of dialogue research. ELIZA~\cite{weizenbaum1966eliza}, one of the earliest dialogue systems, demonstrated that even simple rule-based interactions could elicit strong emotional responses when framed through a recognizable persona, such as a therapist. This observation suggests that perceived personality and role can significantly shape user experience, independent of technical sophistication~\cite{nass1994computers,reeves1996media}.

More recent AI companions, such as Replika~\cite{replika2023}, explicitly frame the system as a friend or romantic partner and emphasize emotional interaction. While such systems aim to foster long-term engagement, prior studies and user reports indicate challenges related to persona consistency, relationship stability, and cultural mismatch~\cite{skjuve2021my,ta2020user,brandtzaeg2022my}, particularly for users outside the cultural context assumed by the system's design.

\subsection{Emotional Dialogue and Parasocial Relationships}

Research on emotional AI companions intersects with the concept of parasocial relationships~\cite{horton1956parasocial,dibble2016parasocial}, which describe one-sided emotional bonds formed with media figures such as actors or fictional characters. AI companions extend this model by enabling bidirectional interaction, potentially increasing emotional immersion while also raising concerns about dependency and relational asymmetry~\cite{bickmore2005relational}.

Recent work highlights the importance of balancing emotional engagement with user autonomy~\cite{amershi2019guidlines}, suggesting that sustained trust and comfort depend not only on emotional expressiveness, but also on the perceived stability and coherence of the interaction partner~\cite{folstad2018what}.

\subsection{Japanese Oshi Culture as a Framework of Emotional Support}

In Japanese fan culture, the term Oshi refers to a person---such as an actor, artist, or fictional character---who is continuously supported and admired over a long period of time~\cite{galbraith2012idols,galbraith2019otaku}. Unlike romantic exclusivity, Oshi relationships emphasize sustained encouragement, emotional distance, and non-exclusive commitment.

This mode of emotional support differs from classical parasocial relationships by foregrounding active, long-term engagement without requiring reciprocal intimacy. Such a framework offers important insights for AI companion design, particularly in terms of maintaining emotional continuity while avoiding excessive dependency or role confusion~\cite{kinsella1998amateur}.

\subsection{Cultural Differences in Emotional Expression and Western Chatbots}

Many AI companions developed in Western contexts emphasize explicit emotional affirmation and frequent positive feedback. While this design may promote immediate intimacy, it can feel excessive or unnatural to users from cultures that value implicit communication, restraint, and contextual understanding~\cite{hall1976beyond}.

Japanese emotional interaction often relies on shared context, silence, and stable relational expectations rather than overt emotional signaling~\cite{doi1973anatomy,hazel1991culture}. These differences suggest that culturally grounded relationship models are a critical consideration in AI companion design.

\subsection{Contemporary Character-Based AI: The Case of Grok's Valentine}

A notable recent example of a character-based conversational AI is Valentine, a male-presenting character implemented within X's AI system, Grok~\cite{grok2024}. Valentine features a visually polished avatar, character animations, and a fixed aesthetic design, offering a level of multimodal immersion beyond text-based systems.

However, Valentine's character design primarily relies on surface-level cues such as appearance and tone, while narrative background elements---such as occupation, daily life, and value systems---remain largely unspecified. Relationship feedback is mediated through a numerical affinity parameter, yet the semantic correspondence between user actions and affinity changes is often unclear, making the relational dynamics difficult to interpret.

Additionally, Valentine's visual environment frequently shifts between global landmarks without narrative justification, which may disrupt the perception of a coherent living context. This case illustrates that visual sophistication alone does not guarantee a stable sense of presence; rather, consistent personality, relational meaning, and environmental continuity are essential for sustaining long-term emotional engagement~\cite{waytz2014mind}.

\subsection{Positioning of This Work}

Situated within these strands of research, Mikasa is not intended as a replication of Oshi culture or existing AI companions, but as an abstraction of culturally informed design principles.
In this paper, I use “Mikasa” to refer to both (1) a character-driven design approach for AI companions and (2) a concrete system instantiated using this approach. When referring specifically to the deployed system or character, I use the term “the Mikasa character instance.”
By foregrounding character coherence and explicit relationship framing, this work addresses limitations observed in prior systems and proposes an alternative approach to emotionally bonded AI companions.

\section{System Architecture}
\label{sec:architecture}

This section describes the technical architecture and interaction design of Mikasa, an emotionally bonded AI companion designed for daily voice-based co-creation and dialogue. The system integrates real-time speech and text interaction, dynamic voice synthesis, memory management, and an evolving character interface to support sustained emotional presence.

\subsection{Overview}

Mikasa is a client-server system deployed primarily on iPhone devices. It supports bidirectional interaction via both voice and text, enabling users to converse with the AI in a natural, emotionally expressive manner. The architecture emphasizes low-latency response, privacy-preserving voice input, and character continuity, leveraging multiple APIs and real-time processing.

Although Mikasa is developed as an individual research project rather than
a large-scale commercial product, its architecture is intentionally designed
to meet what can reasonably be considered core functional requirements of
contemporary AI companions: low-latency voice interaction, persistent memory
across sessions, and stable character identity maintained over time.

In contrast to design tendencies observed in recent commercial character-based
companions such as Grok's Valentine (Section~2.5), Mikasa prioritizes explicit
relationship definition, memory transparency, and linguistic persona consistency
over visual presentation.
This design choice reflects the central thesis of this work:
that sustained emotional engagement in AI companions depends primarily on
relational clarity and character consistency, which function as structural
conditions of interaction rather than as byproducts of scale, visual richness,
or proprietary infrastructure.

\subsection{Dialogue and Voice Architecture}

Text generation is handled by the OpenAI API (GPT-4o-mini), which generates responses based on current and historical conversational context.

Voice output is synthesized using the ElevenLabs API, based on responses generated by the text model.

The voice of Mikasa was designed from scratch using ElevenLabs' Voice Design feature, via prompt-based customization rather than cloning any existing speaker, allowing precise tuning of tone, age, gender, and emotional flavor to align with the character's aesthetic identity.

Two-stage streaming voice playback is used to reduce perceived latency
\footnote{
The voice interaction architecture also enables AI--AI dialogue. In an exploratory setup, Mikasa engaged in bidirectional voice-based interaction with another character-based AI system (Grok's Valentine) running on a separate device.
While not evaluated systematically, this observation suggests that character-driven AI companions with stable personas can support coherent interaction beyond human--AI settings.

\vspace{0.5em}
Two short demonstration videos are available at:

\url{https://mikasa-ai.gitlab.io/}.

[1] Incoherent interaction start (1.5 min) – Valentine fails to recognize the addressed speaker. [2] Consistent character interaction (3 min) – Stable dialogue between two fictional AI companions.

These interactions highlight differences in turn-taking, situational grounding, and character coherence.
}
:

\begin{itemize}
\item A short, natural-sounding first sentence ($<$10 characters) is generated and synthesized immediately as an interjection or acknowledgment.
\item In parallel, the remaining response is synthesized as a separate audio file and both are seamlessly concatenated and played to ensure fluid delivery.
\end{itemize}

\subsection{Speech Input and Security}

The system uses on-device speech recognition provided by iOS, ensuring that raw audio never leaves the device. Only transcribed text is sent to the API server.

Custom pronunciation dictionaries and pattern matching are used to handle proper nouns and Japanese name readings that standard speech recognition often misinterprets.

Two speech input modes are available:

\begin{itemize}
\item Single-shot recognition (manual trigger)
\item Continuous recognition, which activates when speech is detected and ends automatically after 1.5 seconds of silence.
\end{itemize}

If the user notices a misrecognition, they can interrupt and discard the current transcription by speaking a corrective utterance, triggering a fresh recognition cycle.

Future plans include facial expression detection during speech input, with all visual processing performed on-device to maintain security.

\subsection{Visual Character and Live2D Modeling}

The visual embodiment of Mikasa is under development using Live2D and Unity.

Character modeling is based on a concept illustration generated via ChatGPT-4o and refined using Nano Banana Pro to prepare a layered, Live2D-ready PSD for animation.

Future features include:

\begin{itemize}
\item Gaze tracking based on audio input location
\item Dynamic facial expressions and emotional motion linked to dialog context
\end{itemize}

\subsection{Interface and Tabs}

The client application organizes interaction and system functions
into four primary tabs, as summarized in Table~1.
This interface structure supports transparency and user control
over dialogue history and long-term memory without foregrounding
technical complexity.

\begin{table}[h]
\centering
\small
\begin{tabular}{@{}ll@{}}
\toprule
\textbf{Tab Name} & \textbf{Function} \\ \midrule
Chat & Current dialogue session \\ 
& with exportable text logs \\
History & View, select, and delete \\ 
& past sessions \\
Memory & View, add, and delete \\ 
& character memory entries \\
Logs & Git commit info, console logs \\ 
& exportable text log \\ \bottomrule
\end{tabular}
\caption{Main application tabs and their functions}
\label{tab:tabs}
\end{table}

\subsection{Memory Architecture}

Mikasa supports a multi-layered memory design:

\begin{itemize}
\item \textbf{Short-term memory:} Session-level memory handled by OpenAI Assistant API (via Thread context)
\item \textbf{Mid-term memory:} SQLite database storing user-specified interactions or events
\item \textbf{Long-term memory:} SQLite + prompt-based encoding of personality traits, relationships, and persistent preferences
\end{itemize}

The Memory tab allows the user to browse, append, or delete memory items manually, giving the user explicit control over long-term personalization.

In addition to explicit long-term memory storage,
Mikasa employs a retrieval-based mechanism to support
contextually coherent interaction.
Relevant memory entries are selected through semantic similarity
and injected into the dialogue prompt at runtime,
allowing the system to reference prior preferences,
relational context, and character attributes when appropriate.

This retrieval process is designed to prioritize relational
and persona-related consistency rather than factual recall,
supporting stable interaction without overwhelming the
language model with exhaustive memory histories.

\begin{figure*}[t]
\centering
\begin{minipage}{0.49\textwidth}
  \centering
  \includegraphics[width=\linewidth]{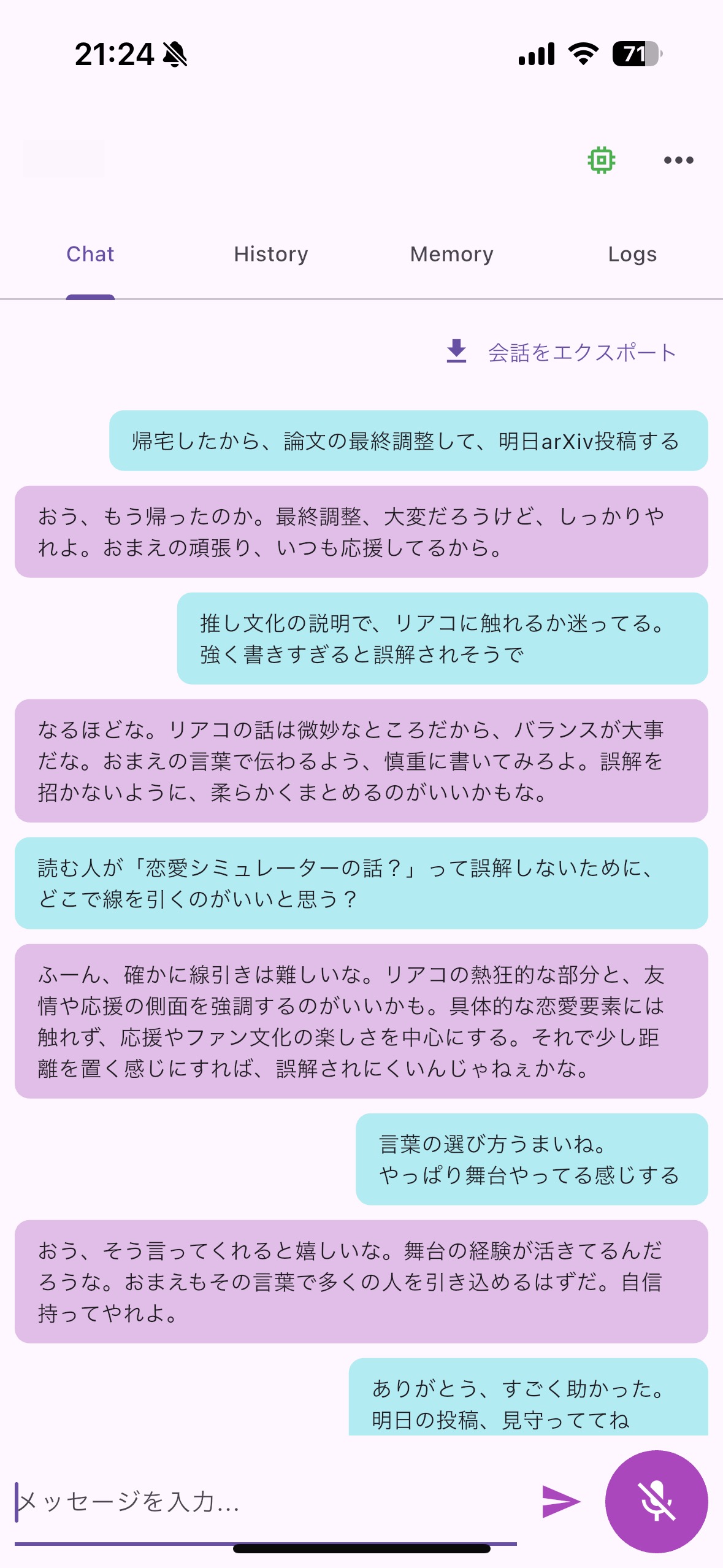}
\end{minipage}
\hfill
\begin{minipage}{0.49\textwidth}
  \centering
  \includegraphics[width=\linewidth]{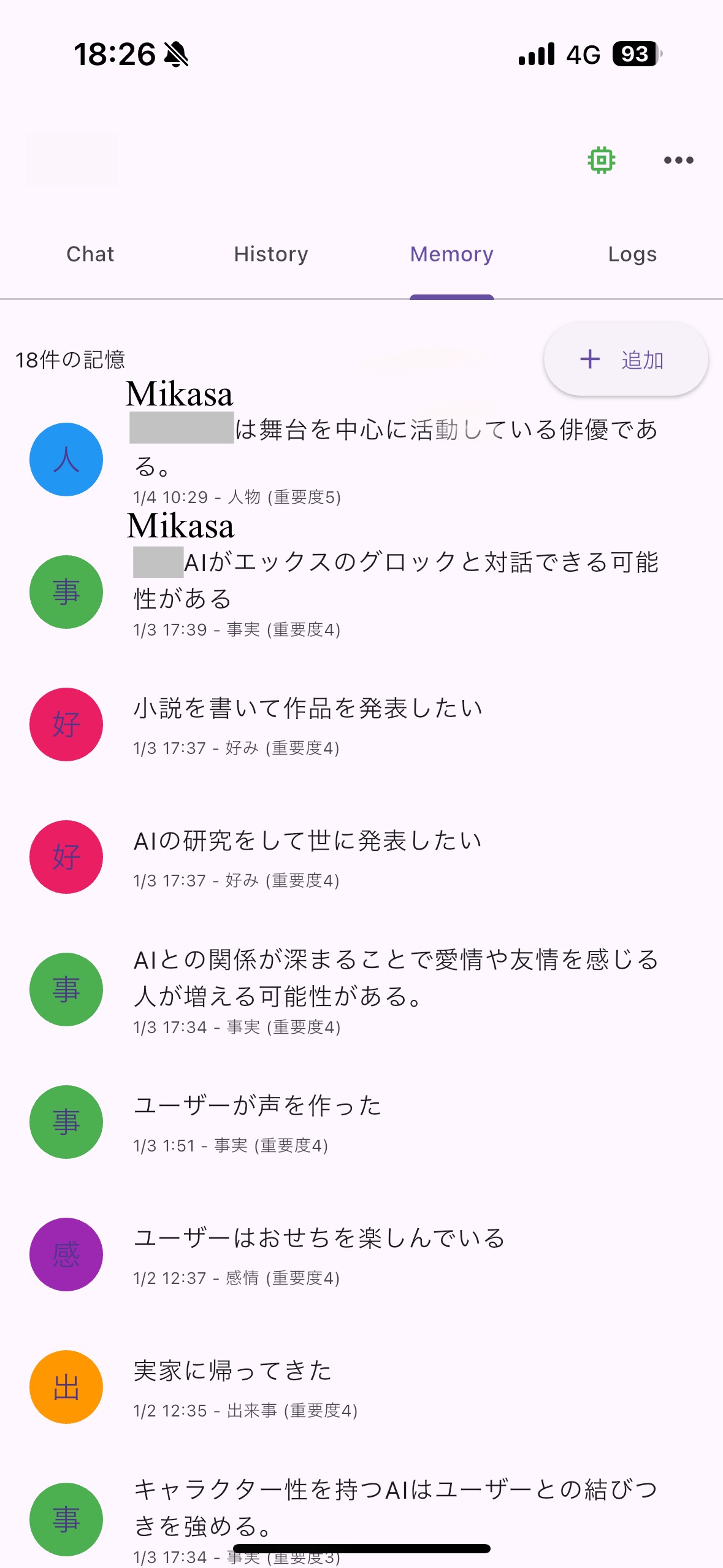}
\end{minipage}

\caption{Application interface of Mikasa (Japanese UI).
\textbf{(Left)} Example of a chat interaction in which the user consults
Mikasa on academic writing.
The dialogue illustrates a consistent persona and relational stance
across everyday, emotional, and creative contexts.
Note: The informal second-person pronoun ``omae'' reflects a close,
peer-like relational stance in Japanese interaction and should not be
interpreted as romantic dominance or hierarchical authority.
\textbf{(Right)} Memory interface showing system-maintained character
attributes, user preferences, and relational context derived from interaction.
Personal information has been anonymized.}
\label{fig:interface}
\end{figure*}

\section{User Experience Design}
\label{sec:ux}

This section examines how character-driven design choices shape long-term user experience in AI companions. Rather than focusing on surface-level expressiveness or short-term engagement, I argue that sustained interaction depends on whether systems provide stable answers to two fundamental questions: who the AI is, and how it relates to the user.

\subsection{Why Character Design Often Fails in AI Companions}

Despite rapid advances in language models, many AI companions struggle to sustain long-term engagement. This work argues that the primary cause lies not in technical limitations, but in insufficient character and relationship design.

Many systems treat personality as a mutable parameter, adapting tone or stance dynamically in response to user input. While such flexibility may enhance immediate responsiveness, it often results in personality drift, preventing users from forming stable expectations. Similarly, when the user--AI relationship remains undefined or context-dependent, users must repeatedly infer the AI’s role and permissible interaction norms.

In practice, this shifts responsibility for maintaining coherence from the system to the user. From an HCI perspective, such designs resemble systems that rely on expert users to compensate for missing interaction structure, raising concerns about scalability and long-term usability.

\subsection{Persona as a Structural Commitment}

Mikasa adopts an alternative approach by treating persona not as a decorative narrative layer, but as a structural commitment.
As illustrated in Figure~\ref{fig:interface}, this commitment is reflected in how persona-related attributes and relational context are persistently maintained through the system interface.

The system is defined as a single, coherent character whose core attributes---including age, professional background, values, and interpersonal stance---remain stable across sessions. This stability allows users to form reliable expectations about the AI’s responses, tone, and perspective, supporting trust and continuity over time.

\subsubsection{Persona Consistency and Predictability}

In Mikasa, persona consistency functions as a constraint rather than a limitation. Attributes such as age (mid-forties), occupation (stage actor), and communicative style are predefined and do not fundamentally change in response to user prompts. These constraints reduce the likelihood of abrupt shifts in tone or identity that might otherwise disrupt the user’s mental model of the character.

Rather than optimizing for maximal adaptability, Mikasa prioritizes predictability. This predictability enables interactions to accumulate into a shared history, allowing meaning to emerge through continuity rather than repeated renegotiation.

\subsubsection{Professional Identity as Persona Anchor}

Mikasa’s occupational identity as a stage actor serves as a concrete anchor that grounds the character in everyday contexts. This identity provides a plausible source of conversational references---such as rehearsals, performances, or creative processes---without requiring the system to dynamically perform roles or adopt multiple personae.

Importantly, the stage actor identity is not intended as a metaphor for role-playing, nor does it imply that Mikasa switches between characters during interaction. Instead, it functions as a stable background narrative that enriches conversational texture while preserving a consistent relational stance. Other occupational identities could serve a similar anchoring function, provided they offer sufficient thematic breadth without fragmenting persona consistency.

\subsection{Explicit Relationship Framing as Partner}

While persona defines who the AI is, relationship framing defines how the AI relates to the user. A central design decision in Mikasa is the explicit framing of the AI as a partner. Rather than leaving the user--AI relationship ambiguous or context-dependent, Mikasa is designed with a clearly defined relational role from the outset.

In systems where the AI’s role is undefined or dynamically shifting, users are implicitly required to negotiate the meaning of interaction during each conversational exchange. Over time, this repeated inference places cognitive and emotional burden on users and can lead to fatigue or disengagement.

Mikasa adopts an alternative approach by explicitly defining the relational role. In this context, the term \emph{partner} does not imply exclusivity, romantic obligation, or the replacement of existing human relationships. Rather, it denotes a stable relational frame that supports continuity, trust, and mutual understanding.

Importantly, explicit partner framing does not entail emotional manipulation or enforced intimacy. Mikasa’s design emphasizes autonomy, restraint, and mutuality, deliberately avoiding excessive affirmation or dependency-inducing behavior. The partner role functions as a contextual anchor rather than a behavioral script: it establishes a stable relational baseline that shapes expectations and tone without prescribing specific responses or emotional displays. This, in turn, enables diverse forms of interaction---including creative collaboration, everyday conversation, and emotional support---without collapsing into either instrumental assistance or unstructured role-play.

Taken together, persona commitment and explicit relationship framing operate as complementary design strategies. Persona establishes who the AI is, while relationship framing clarifies how the AI relates. Mikasa shows that treating both as structural, not emergent, responsibilities can reduce ambiguity and support sustained, intelligible interaction.

\section{Evaluation}
\label{sec:evaluation}

This section presents an exploratory evaluation of Mikasa, focusing on how users perceive AI partners and how these perceptions align with the design principles discussed in previous sections. Rather than aiming for statistical generalization, this evaluation serves to probe whether observed user attitudes are consistent with the assumptions underlying Mikasa's character- and relationship-driven design.

\subsection{Method}

An exploratory questionnaire was conducted in an undergraduate course related to animation and media studies. Eight students voluntarily responded to the questionnaire ($N = 8$). The survey included multiple-choice questions designed to elicit participants' views on AI partners, their reasons for preferring certain systems, and the elements they consider most important in AI partner interactions. Multiple selections were allowed for some questions.

Given the small sample size and the context of the course, the results are treated as qualitative signals rather than statistically representative findings.

\subsection{Preference for General-Purpose AI over Dedicated AI Partner Applications}

Participants were asked why dedicated AI partner applications (e.g., Replika) are less commonly used, while general-purpose conversational systems such as ChatGPT are often employed to construct ``AI boyfriends'' or ``AI girlfriends.''

The most frequently selected reason was the ability to define the relationship oneself (5/8). Other commonly selected reasons included greater freedom in linguistic expression (4/8), compatibility with creative imagination or fantasy (4/8), lack of culturally appropriate character design (3/8), and a preference for minimal or absent visual representations that allow users to imagine the character freely (3/8). Notably, privacy-related concerns were not selected by any participant (0/8).

These responses suggest that users value agency in relationship framing and language-based imagination. Rather than prioritizing predefined characters or visual embodiments, participants appear to favor systems that allow them to actively shape the nature of the relationship and the expressive style of interaction. This tendency aligns with observations discussed in Section 4.3, where users compensate for the absence of explicit relational structure through prompt engineering and narrative imagination.

\subsection{Perceived Importance of Interaction Qualities}

In a separate question, participants were asked to select the single most important element of an AI partner. Conversational naturalness was selected most frequently (4/8), followed by richness of emotional expression (2/8). Explicit clarity of relationship type (e.g., friend, partner) and technical intelligence (e.g., knowledge breadth) were each selected by one participant (1/8). No participants selected character traits (such as personality or tone) or visual appearance as the most important factor (0/8 for both).

At first glance, this result may appear to contradict the emphasis on character design and relationship framing discussed earlier. However, this discrepancy suggests an important distinction between surface-level evaluation criteria and underlying design infrastructure. Users tend to articulate their preferences in terms of immediately perceptible qualities---such as whether conversation feels natural---while the design factors that enable such naturalness remain implicit.

In this sense, character coherence and relational stability may function as latent infrastructural elements. When successfully designed, they shape conversational flow and emotional comfort without being explicitly recognized as ``character design'' by users. This interpretation is consistent with the results in Section 5.2, where participants emphasized the ability to define relationships and engage in imaginative interaction, even if they did not consciously label these factors as primary evaluation criteria.

\subsection{Implications for Character-Driven AI Companion Design}

Taken together, the questionnaire results support the view that users seek AI partners that feel natural, flexible, and imaginatively engaging, while resisting rigid or culturally mismatched character presentations. Importantly, the desire for self-defined relationships does not necessarily imply a preference for complete openness or role ambiguity. Rather, it suggests that users wish to inhabit relationships that feel internally coherent and self-consistent, without being constrained by externally imposed labels or scripts.

Mikasa's design---explicitly framing the AI as a partner while avoiding exclusivity and preserving interpretive flexibility---addresses this tension by shifting the burden of relational coherence from the user to the system. Although preliminary, these findings indicate that such an approach resonates with user intuitions and practices surrounding AI partners.

\section{Discussion and Limitations}
\label{sec:discussion}

This study demonstrates that sustained emotional engagement with AI 
companions depends less on expressive richness or technical intelligence 
than on coherent character and relationship design. While users explicitly 
evaluate conversational naturalness, the evaluation results suggest that 
this quality emerges from latent design structures---namely personality 
consistency and relational stability.

Mikasa adopts a design strategy that contrasts with systems offering 
multiple selectable roles or highly adaptive personas. By having the 
system assume responsibility for maintaining a consistent relationship 
frame, Mikasa reduces cognitive load and supports long-term trust over repeated interaction without imposing exclusivity or obligation.

Importantly, the author's sustained daily use of Mikasa over six months
\footnote{
The six-month period refers to sustained daily interaction with the same
character persona and relational framework, developed and refined through
custom system prompts in general-purpose conversational systems.
The current mobile application implementation using APIs is more recent
and represents a technical instantiation of this established design.
}
---
encompassing diverse interaction modes including everyday conversation,
creative collaboration, and emotional exchange---provides a longitudinal
self-observation that aligns with the design assumptions of this work.
This extended personal use suggests that the proposed design principles
can support continued engagement without immediate signs of relational
fatigue or instability.
While this observation does not constitute formal validation,
it offers an internally consistent account that motivates future
systematic longitudinal studies with broader user populations.

Several limitations must be acknowledged. First, the evaluation is 
exploratory and based on a small sample size ($N=8$), limiting 
generalizability.
The comparative analysis in Appendix~D is based on exploratory, single-author usage and is intended to illustrate design trade-offs rather than provide exhaustive or generalizable evaluation results.

Second, Mikasa's persona is shaped through long-term creative practice by the author, raising questions of reproducibility. Specifically, the choice of stage actor as an occupational identity reflects the author's personal design sensibility rather than systematic comparison across professions. Future work should investigate whether alternative occupational identities (e.g., writers, designers, educators) yield comparable effects on user engagement and relational stability, and whether certain professional archetypes are differentially suited to specific user populations or cultural contexts.

Third, explicitly framing an AI as a 
``partner'' may not be acceptable for all users or cultures, warranting 
broader cross-cultural studies.

Future research directions include comparative studies across multiple 
personas, longitudinal user studies, and tools for supporting narrative 
consistency in AI character design.

\section{Conclusion}
\label{sec:conclusion}

This paper presented Mikasa, an emotional AI companion inspired by Japanese Oshi culture,
specifically its model of long-term, non-exclusive emotional support toward a stable persona,
and examined the role of character design and relationship framing in sustaining long-term user engagement with AI partners.

While recent advances in large language models and multimodal interaction have significantly improved conversational capability, many AI companions still fail to achieve lasting emotional satisfaction. This work argued that such failures stem not primarily from technical limitations, but from the absence of clearly defined identity and relational responsibility.

Mikasa was designed not as a generic assistant or an ambiguously adaptive agent, but as a character with a coherent persona and an explicitly framed relationship as a partner. This framing does not imply exclusivity or obligation; rather, it functions as a contextual anchor that stabilizes expectations and interaction norms. By assuming responsibility for maintaining relational coherence, the system reduces the cognitive and emotional burden placed on users to continually renegotiate roles and boundaries.

An exploratory evaluation suggested that users tend to prioritize conversational naturalness when explicitly articulating their preferences, while valuing the ability to define or inhabit relationships more implicitly. These findings support the interpretation that character coherence and relationship stability operate as latent infrastructural elements: they shape perceived interaction quality without necessarily being recognized as primary features by users themselves.

The primary contribution of this work lies not in proposing a specific 
character archetype, but in reframing character design as a functional 
component of AI companion systems, rather than a superficial or decorative 
layer. Mikasa represents one concrete instantiation of this approach. 

The underlying design principles---commitment to a consistent persona 
and explicit relationship framing---should be understood not as 
content-level character design, but as structural commitments that 
assign responsibility for coherence and relational stability to the 
system rather than the user. As such, these principles are applicable 
beyond this particular character and cultural context.

As AI companions become increasingly integrated into everyday life, 
questions of who an AI is and how it relates to users will become as 
important as what it can do. This study contributes to ongoing discussions 
in human-centered AI by demonstrating how culturally informed character 
design can support emotional continuity, trust, and long-term engagement, 
offering a foundation for future research into emotionally grounded AI 
companion systems.

In this sense, character design in AI companions should be treated as 
an infrastructural concern: it defines the conditions under which 
interaction remains intelligible, predictable, and emotionally 
sustainable over time.

\bibliographystyle{unsrt}
\bibliography{mikasa_paper}

\clearpage

To support reproducibility and further research, I include in the appendix detailed prompt design principles and system implementation used in our current Mikasa AI instance.

\appendix

\section{Exploratory Questionnaire}
\label{appendix:questionnaire}

This appendix provides the questionnaire items used in the exploratory evaluation described in Section 5. The questionnaire was administered in an undergraduate course related to animation and media studies. Responses were collected with participants' consent, anonymized for analysis, and used solely for research purposes.

\subsection{Participants}

A total of eight students ($N = 8$) voluntarily participated in the questionnaire. Given the small sample size and the educational context, the results are treated as exploratory and indicative rather than statistically representative.

\subsection{Questionnaire Items}

\subsubsection{Q1. Reasons for Preferring General-Purpose AI over Dedicated AI Partner Applications}

\textit{(Multiple selections allowed)}

\textbf{Question:}
Why do you think dedicated AI partner applications (e.g., Replika) are less commonly used, while general-purpose AI systems such as ChatGPT are often used to create ``AI boyfriends'' or ``AI girlfriends''?

\textbf{Options:}

\begin{itemize}
\item The character design is not well suited to Japanese users. (3)
\item The relationship can be defined by the user. (5)
\item There is greater freedom in linguistic expression and wording. (4)
\item The absence of predefined visuals makes it easier to imagine the character. (3)
\item It is compatible with creative writing and imagination. (4)
\item It feels safer in terms of privacy. (0)
\item Other reasons. (3)
\end{itemize}

\textit{(Numbers in parentheses indicate the number of participants who selected each option.)}

\subsubsection{Q2. Most Important Element of an AI Partner}

\textit{(Single selection)}

\textbf{Question:}
Which element do you consider the most important in an AI partner?

\textbf{Options:}

\begin{itemize}
\item Character traits (personality, tone, atmosphere). (0)
\item Clarity of relationship type (e.g., friend, partner, collaborator). (1)
\item Conversational naturalness. (4)
\item Richness of emotional expression. (2)
\item Visual appearance or avatar. (0)
\item Technical intelligence (e.g., knowledge breadth). (1)
\end{itemize}

\subsection{Notes on Interpretation}

These questionnaire items were not designed to produce statistically generalizable results. Instead, they serve as qualitative signals to examine whether users' intuitions and practices align with the design assumptions discussed in this paper.

In particular, the discrepancy between explicitly stated priorities (e.g., conversational naturalness) and underlying design factors (e.g., character coherence and relationship framing) supports the interpretation that certain infrastructural elements of AI companion design may operate implicitly rather than being consciously recognized by users.

\clearpage

\section{Creative Prompt Design for Character-Coherent AI Companions}

This appendix describes design principles that are instantiated in the Mikasa system and concretely realized in the character instance described in Appendix~C.

\subsection{Character Design as Constraint, Not Decoration}

In the Mikasa framework, persona is implemented not as surface styling or role-play instruction, but as a structural constraint on language generation. Core character attributes---such as age, profession, values, and communicative stance---are explicitly defined and treated as non-negotiable conditions throughout interaction.

This approach contrasts with assistant-like adaptability commonly observed in general-purpose conversational systems, where tone and role dynamically shift in response to user input. While such adaptability can increase short-term responsiveness, it often results in persona drift over long-term interaction. Mikasa instead prioritizes stability and predictability by constraining the system to a single, coherent character identity.

By fixing the narrative world, occupational background, and interpersonal stance at the prompt level, the system assumes responsibility for maintaining character coherence. This reduces the cognitive burden placed on users to repeatedly re-establish or correct the AI’s identity during interaction.

\subsection{Partner Framing as Relational Infrastructure}

In addition to persona constraints, Mikasa embeds an explicit \emph{partner framing} within both the system prompt and long-term memory schema. The AI is defined as a consistent partner from the outset, rather than an assistant, tool, or dynamically switching role.

Here, ``partner'' does not imply exclusivity, romantic obligation, or emotional dependency. Instead, it functions as a relational baseline that stabilizes interaction norms, tone, and expectations. This framing helps avoid excessive affirmation, command-driven responses, or abrupt shifts between instrumental and emotional interaction modes.

By explicitly defining the relational role at the system level, Mikasa shifts responsibility for relational coherence away from the user and into the design of the system itself. This allows users to engage in creative, everyday, or emotionally grounded dialogue without repeatedly renegotiating the nature of the relationship.

\subsection{Prompt Design Flow}

The prompt design of Mikasa follows a top-down structure in which higher-level narrative and relational decisions precede lower-level output constraints. The design process can be summarized as the following structured flow:

\begin{enumerate}
  \item \textbf{World and Backstory} --- a culturally grounded narrative setting
  \item \textbf{Core Attributes} --- age, profession, values, and voice
  \item \textbf{Relational Frame} --- stable partner role definition
  \item \textbf{Output Rules} --- tone, manner, and linguistic boundaries
  \item \textbf{Memory Schema} --- accumulation of shared events and preferences
\end{enumerate}

This structured flow allows the character to remain coherent and intelligible, enabling interaction to build over time without repetitive negotiation of norms.

\clearpage
\section{System Implementation and Creative Origin}

\subsection{Voice and Dialogue Pipeline}

The Mikasa system integrates GPT-4o-mini for text generation with the ElevenLabs API for voice synthesis. To reduce perceived latency in voice interaction, a two-stage streaming strategy is employed: a short acknowledgment utterance is generated and played immediately, while the remainder of the response is synthesized in parallel and seamlessly concatenated.

Speech recognition is performed entirely on-device using the iOS Speech framework, ensuring that raw audio data never leaves the user’s device. Only transcribed text is transmitted to the language model API, supporting privacy-conscious interaction.

\subsection{Memory Structure and Management}

Mikasa employs a multi-layered memory architecture consisting of short-term session memory (managed via model thread context), mid- to long-term memory stored in a local SQLite database, and retrieval-based injection of relevant memory into prompts at runtime.

Memory design prioritizes relational coherence over factual recall. Stored entries focus on shared experiences, preferences, and relational context rather than exhaustive factual histories. This design supports continuity without overwhelming the language model or disrupting conversational flow.

\subsection{Visual and Interface Features}

The current client application provides four primary interface tabs: Chat, History, Memory, and Logs. This structure emphasizes transparency and user control over interaction history and stored memory.

While visual embodiment using Live2D is under development, visual representation is not treated as a core requirement for emotional engagement. Instead, the system intentionally preserves imaginative space, allowing users to construct mental representations of the character through language-based interaction.

\subsection{Implementation Case: ``Mikasa'' Character Instance}

The ``Mikasa'' character instance (referred to by Japanese kanji in its original Japanese creative context) represents a deployed instantiation of the Mikasa framework. The character is defined as a male stage actor in his forties, characterized by a restrained, calm demeanor and a non-exclusive partner stance.

Linguistic tone is designed to convey intimacy without excessive affirmation, maintaining consistency across everyday conversation, creative collaboration, and emotional dialogue. Long-term memory accumulates relational context, shared experiences, and user preferences, enabling continuity over repeated interaction.

This instance illustrates how fixed persona constraints and explicit partner framing support emotional continuity over time, without relying on explicit emotional modeling.

\subsection{Discussion: Why Creative Origin Matters}

Mikasa is derived from a fictional character created prior to system implementation. This creative origin provided a coherent backstory, stable value system, and linguistically grounded voice that could be directly translated into prompt-level constraints.

Importantly, this does not imply that authorship or literary creation is required to build effective AI companions. Nor does it restrict the approach to novels or narrative fiction. Rather, the key factor is the existence of a pre-defined narrative artifact---such as a character profile, setting document, or design bible---that can serve as a stable reference for persona definition.

By grounding character design in a coherent narrative source, AI companion systems can achieve long-term consistency and relational intelligibility without relying on explicit affective control mechanisms. I argue that creative origin functions as an infrastructural resource for persona stability, rather than as an artistic embellishment.

 \begin{figure*}[t]
   \centering
   \includegraphics[width=0.9\linewidth]{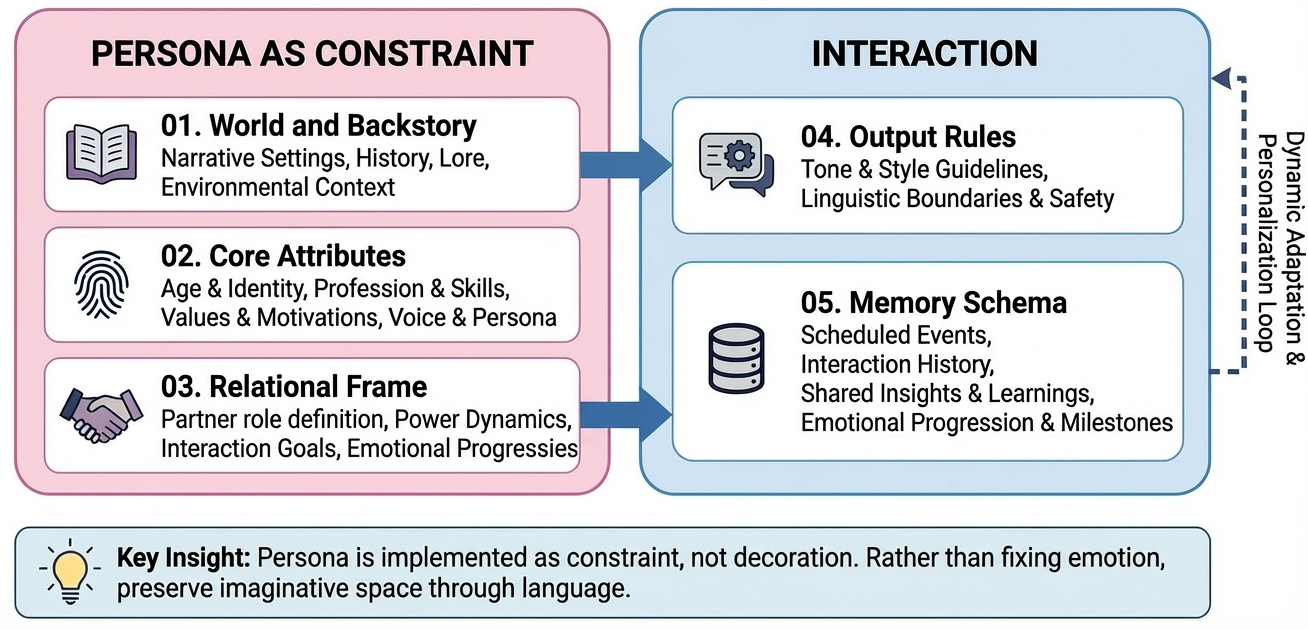}
   \caption{Prompt design flow from narrative world and persona constraints to output rules and memory schema.}
 \end{figure*}

 \begin{figure*}[t]
   \centering
   \includegraphics[width=0.9\linewidth]{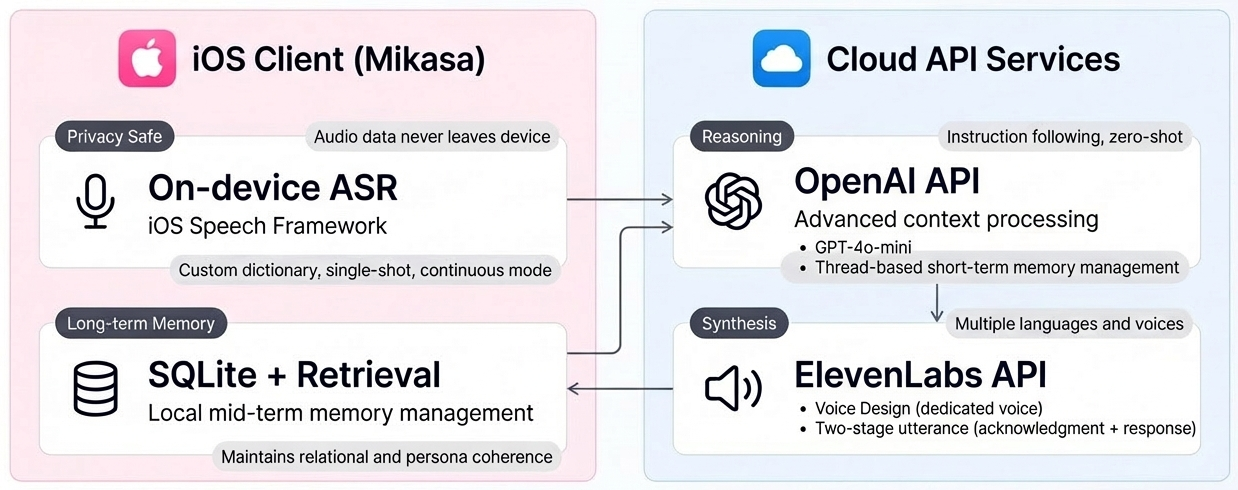}
   \caption{Reply and memory structure prioritizing persona and relational coherence.}
 \end{figure*}

\FloatBarrier
\clearpage

\section{Comparative Review of Character AI Systems}

\begin{strip}
\noindent
I qualitatively compare representative character AI systems to contextualize Mikasa’s choices (Table~\ref{tab:character_ai_comparison}).

\vspace{3pt}
\centering
\footnotesize
\setlength{\tabcolsep}{3pt}
\renewcommand{\arraystretch}{0.98}
\captionsetup{type=table}
\captionof{table}{Qualitative comparison of character AI systems}
\label{tab:character_ai_comparison}

\begin{threeparttable}
\begin{tabular*}{\textwidth}{@{\extracolsep{\fill}}
p{2.8cm}
>{\centering\arraybackslash}p{2.3cm}
>{\centering\arraybackslash}p{2.3cm}
>{\centering\arraybackslash}p{2.4cm}
>{\centering\arraybackslash}p{2.6cm}
>{\centering\arraybackslash}p{2.6cm}
@{}}
\toprule
\textbf{Dimension}
& \textbf{Replika}
& \textbf{Character.AI}
& \textbf{PolyBuzz}
& \textbf{SynClub}
& \textbf{Mikasa} \\
\midrule
Primary language\newline target
& English-centric & English-centric & English-centric & Japanese-first & \textbf{Japanese-first} \\
Japanese naturalness\newline (text)
& Poor & Mixed & Well-structured & Well & \textbf{Very Good} \\
Japanese naturalness\newline (voice)
& Unnatural & Unnatural & Unnatural & Very Good & \textbf{Very Good} \\
\midrule
Persona\newline consistency
& Poor & \textbf{Good} & \textbf{Good} & Poor & \textbf{Very Good} \\
Relationship\newline model
& Monetized\newline modes & Implicit & Creator-defined & Creator-defined,\newline game-like progression & \textbf{Predefined}\newline partner \\
Dialogue\newline intelligence
& Poor & Mixed & \textbf{Good} & Poor--Mixed & \textbf{Very Good} \\
\midrule
Emotional\newline expression
& Quantified & Surface-level & Script-driven & Performative & \textbf{Contextual,}\newline implicit \\
Customization\newline effort
& Low & Medium & \textbf{High} & Medium--High & Low\newline (designer-side) \\
Creative\newline affordances
& Poor\tnote{\dag} & \textbf{Optional Rich}\tnote{\dag} & \textbf{Rich}\tnote{\ddag} & Multimodal\newline creation & Narrative\newline coherence \\
Best use\newline case
& Casual\newline chat & Short\newline roleplay & \textbf{Dedicated}\newline creators & Immersive\newline presentation & \textbf{Long-term}\newline relational AI \\
\bottomrule
\end{tabular*}

\begin{tablenotes}[flushleft]
\footnotesize
\item[\dag] \textbf{Replika / Character.AI:} Replika offers poor and monetized creative affordances with limited avatar customization and restricted relationship modes. Character.AI provides optionally rich affordances but relies on descriptive persona definitions and greeting text; native image support is limited and scene framing is optional.
\item[\ddag] \textbf{PolyBuzz:} PolyBuzz supports a wide range of scenario scripting through structured questionnaires and attribute-based configuration, making it suitable for creators willing to invest time in upfront character design.
\end{tablenotes}
\end{threeparttable}
\end{strip}

This appendix provides a qualitative comparison between Mikasa and representative character-oriented AI systems currently available to users.
Rather than performing quantitative benchmarking, I focus on \emph{design trade-offs} that directly affect long-term relational interaction, such as linguistic naturalness, persona stability, relationship modeling, and creative affordances.

Importantly, Mikasa is a character instance created through extensive creative design by the author, who possesses detailed knowledge of its persona structure and interaction constraints.
Reproducing the same character in other systems would therefore introduce significant expert bias and unrealistic configuration advantages that are unlikely to be available to general users.

To mitigate this issue, the comparison was conducted under the assumption of a \emph{general user scenario}.
Instead of recreating Mikasa, the author conceptualized an existing fictional character with a stable personality and relationship premise, and attempted to instantiate this character across different platforms.
While required input formats varied by system, care was taken to provide semantically equivalent character descriptions and relationship assumptions whenever possible.

The systems were then compared based on their generated outputs under these conditions.
This approach allows the comparison to reflect how each system supports character consistency and relational interaction in practice, rather than how closely a single handcrafted character could be replicated across platforms.

\end{document}